# The Efficiency Limit of $CH_3NH_3PbI_3$ Perovskite Solar Cells

Wei E.I. Sha[1,2], Xingang Ren[1], Luzhou Chen[1], and Wallace C.H. Choy[1, a]

1. Department of Electrical and Electronic Engineering, The University of Hong Kong, Pokfulam Road, Hong Kong, China

2. The University of Hong Kong Shenzhen Institute of Research and Innovation (HKU-SIRI), Shenzhen, 518057, China.

[a] Electronic mail: chchoy@eee.hku.hk; wsha@eee.hku.hk

**Abstract**

With the consideration of photon recycling effect, the efficiency limit of methylammonium lead iodide ($CH_3NH_3PbI_3$) perovskite solar cells is predicted by a detailed balance model. To obtain convincing predictions, both AM 1.5 spectrum of Sun and experimentally measured complex refractive index of perovskite material are employed in the detailed balance model. The roles of light trapping and angular restriction in improving the maximal output power of thin-film perovskite solar cells are also clarified. The efficiency limit of perovskite cells (without the angular restriction) is about 31%, which approaches to Shockley-Queisser limit (33%) achievable by gallium arsenide (GaAs) cells. Moreover, the Shockley-Queisser limit could be reached with a 200 nm-thick perovskite solar cell, through integrating a wavelength-dependent angular-restriction design with a textured light-trapping structure. Additionally, the influence of the trap-assisted nonradiative recombination on the device efficiency is investigated. The work is fundamentally important to high-performance perovskite photovoltaics.

Recently, organic-inorganic $(CH_3NH_3)PbX_3$ perovskite semiconductors (where Pb is lead and X can be iodine, bromine or chlorine) have been studied and applied in fabricating thin-film photovoltaic devices because of their low-cost and potential for high efficiency. Due to merits of direct bandgap [1-4], high and balanced carrier mobility [5], long electron-hole diffusion length [6-8], and low nonradiative Auger recombination [9], power conversion efficiency (PCE) of 20.1% has been certified for perovskite solar cells [10]. Consequently, it is the right time to predict the efficiency limit of perovskite solar cells, which provides important physical insights to further upgrade the PCE of practical perovskite photovoltaic devices.

According to the detailed balance model by Shockley and Queisser [11-26], the maximal output power of a solar cell can be achieved if the following set of hypotheses are fulfilled:

(1) Carrier populations obey Maxwell-Boltzman statistics. Particularly, the quasi-Fermi levels of electrons and holes are uniformly split through the cell and the split equals the applied voltage. The assumption is reasonable if mobility of photocarriers (electrons and holes) are sufficiently large. Regarding perovskite materials, charge carrier mobility as high as 10 $cm^2\,V^{-1}\,s^{-1}$ has been observed [5].

(2) Radiative band-to-band (bimolecular) recombination mechanism is the only one existing. Nonradiative recombination, such as Auger recombination, trap (defect) assisted recombination, etc, are ignorable. Different from silicon with an indirect bandgap, perovskite material has a direct band gap. Therefore, Auger recombination is sufficiently suppressed, which has been verified in recent experimental results [9]. Moreover, light emission from perovskite solar cells is dominated by a sharp band-to-band transition that has a radiative efficiency much higher than that of organic solar cells [27].

(3) Internal conversion efficiency reaches 100%. When one photon is absorbed, it produces one electron-hole pair; and when one electron-hole pair recombines, it produces one photon.

(4) Photon recycling effect [28-33] occurs in the cell. Although a photon will be created by one electron-hole pair recombination during the radiative recombination process, the photon can be reabsorbed at a different spatial location in the cell, which creates a new electron-hole pair. Designs of light trapping and angular restriction [29, 31, 32] can improve the photon reabsorption process and thus maximize solar cell efficiency.

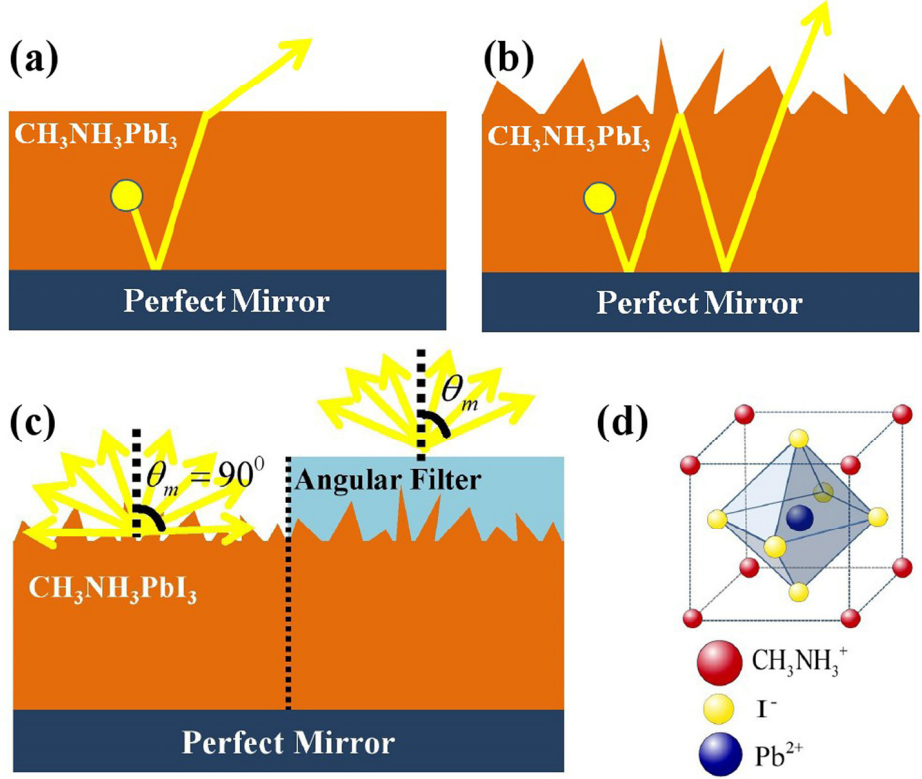

**FIG. 1.** (Color online) $CH_3NH_3PbI_3$ solar cells with different light-trapping and angular-restriction structures. (a) a planar front surface with a perfectly reflecting mirror on the rear surface; (b) a randomly textured front surface and a perfectly reflecting mirror on the rear surface; (c) textured solar cells with and without the angular-restriction filter; (d) a crystal structure of $CH_3NH_3PbI_3$.

In the detailed balance model, the photocurrent is calculated as the difference between the photons absorbed and the photons leaving the cell.

$$J(V) = J_e(V) - J_p \qquad (1)$$

where $V$ is the applied voltage of the solar cell system. $J_e$ describes current density loss due to radiative emission, and $J_p$ is the photocurrent (photogenerated current) due to absorption of incident sunlight.

The photogenerated current is given by

$$J_p = q \int_0^\infty a(\lambda, W) \frac{\Gamma(\lambda)\lambda}{hc_0} d\lambda \qquad (2)$$

where $c_0$ is the speed of light in free space, $\Gamma$ is the global AM1.5G spectrum of Sun (W·m$^{-2}$·nm$^{-1}$), $\lambda$ is the wavelength and $q$ is the elementary charge. The absorptivity $a$ depends on light-trapping and angular-restriction designs. For solar cells having a planar front surface with a perfectly reflecting mirror on the rear [See **Fig. 1(a)**], we have

$$a(\lambda, W) = 1 - \exp[-2\alpha(\lambda)W] \qquad (3)$$

where $W$ is the cell thickness. The constant 2 is due to doubled optical path under vertical incidence condition; and reflection from the cell can be ignored by using a perfect antireflective coating. $\alpha = 2k_0 k_i$ is the absorption coefficient of active (semiconductor) material, $k_0$ is the wavenumber of free space, and $k_i$ is the imaginary part of complex refractive index of active material. For solar cells having a randomly textured front surface with a perfectly reflecting mirror on the rear [See **Fig. 1(b)**], the absorptivity can be written as [13, 34]

$$a(\lambda, W, \theta_m) = \frac{\alpha(\lambda)}{\alpha(\lambda) + \frac{\sin^2 \theta_m(\lambda)}{4n_r(\lambda)^2 W}} \qquad (4)$$

where $n_r$ is the real part of complex refractive index of the active material. $\theta_m$ is the maximum angle of emission (with respect to the normal of the front surface), i.e. light escapes out of the solar cell within a cone with a solid angle of $\theta_m$. The maximum angle of emission can be reduced by using an angular-restriction filter as shown in **Fig. 1(c)**.

According to Maxwell-Boltzmann statistics, the radiative current density is expressed as

$$J_e(V) = J_0 \left[ \exp\left(\frac{qV}{k_B T}\right) - 1 \right] \qquad (5)$$

where $k_B$ is the Boltzmann constant and $T$ is the Kelvin temperature. The above is the same as Shockley's diode equation [28] but the dark current $J_0$ should be represented with the aid of the black-body radiation law. Solar cells absorb photon energy and randomly generate electron-hole pairs, which are the physical original of dark current. Meanwhile, photons are emitted immediately after electrons and holes recombine with each other. Under thermal equilibrium condition, the quantity of absorbed photons should be balanced with that of emitted photons. The thermal spectral radiance of a cell, i.e. power emitted per area per wavelength per solid angle, is given by

$$S(\lambda) = \frac{2hc_0^2}{\lambda^5} \frac{1}{\exp\left(\frac{hc_0}{\lambda k_B T}\right) - 1} \qquad (6)$$

From Eq. (6), the black-body (thermal) emission spectrum of the cell, i.e. power emitted per area per wavelength, can be calculated by the surface integral in spherical coordinates

$$\Gamma_0(\lambda) = \int_0^{2\pi} d\varphi \int_0^{\theta_m} S(\lambda)\cos(\theta)\sin(\theta)d\theta = \pi \sin^2\theta_m(\lambda) S(\lambda) \qquad (7)$$

where $\cos(\theta)$ is due to the angle of emission with respect to the normal of the front surface. Finally, the dark current is formulated as

$$J_0 = q\int_0^\infty a(\lambda, W)\frac{\Gamma_0(\lambda)\lambda}{hc_0}d\lambda \qquad (8)$$

The dark current of Eq. (8) shows a similar mathematical expression to the photocurrent of Eq. (2) except the sun spectrum $\Gamma$ should be replaced by the black-body (thermal) emission spectrum of solar cell $\Gamma_0$. Since Boltzmann approximation is adopted for the term $J_0 \exp\left[qV/(k_B T)\right]$ in Eq. (5), the exponential term $\exp\left[hc_0/(\lambda k_B T)\right]$ in Eq. (6) is replaced by $\exp\left[(hc_0/\lambda - qV)/(k_B T)\right]$ for an exact calculation of the term.

**Figure 1(d)** shows a crystal structure of $CH_3NH_3PbI_3$ perovskite. The complex refractive index of $CH_3NH_3PbI_3$ perovskite material is described by Forouhi-Bloomer model, which is in well agreement with experimental results measured via the angle spectroscopic ellipsometry and spectrophotometry [35]. **Figure S1** shows the complex refractive index of $CH_3NH_3PbI_3$ materials by the Forouhi-Bloomer model (See Supplementary Material [36]). In view of spectral absorbing regions of perovskite materials, the effective wavelengths in integrals of Eqs. (2) and (8) are ranged from 280 nm to 1000 nm. (In other words, lower and upper limits of the integrals are 280 nm and 1000 nm, respectively.)

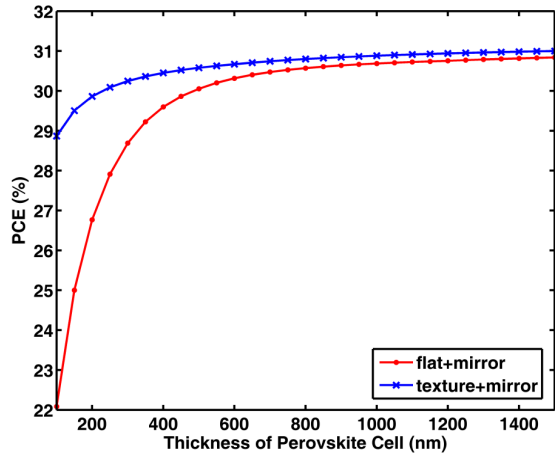

**FIG. 2.** (Color online) PCE of $CH_3NH_3PbI_3$ perovskite solar cells as a function of cell thickness. Flat and textured cells adopt different light-trapping structures as depicted in **Figs. 1(a) and 1(b)**. No angular restriction is adopted, i.e. the maximum emission angle $\theta_m = 90°$.

**Table 1.** Electrical characteristics for $CH_3NH_3PbI_3$ perovskite solar cells. Flat and textured cells adopt different light-trapping structures as depicted in **Figs. 1(a)** and **1(b)**, respectively. $J_{sc}$, $V_{oc}$, FF, and PCE are the short notations of short-circuit current, open-circuit voltage, fill factor, and power conversion efficiency, respectively.

|  | flat and perfect mirror | | | textured and perfect mirror | | |
|---|---|---|---|---|---|---|
| Thickness (nm) | 200 | 500 | 1000 | 200 | 500 | 1000 |
| $V_{oc}$ (V) | 1.325 | 1.315 | 1.305 | 1.300 | 1.295 | 1.290 |
| $J_{sc}$ (mA/cm$^2$) | 22.27 | 25.27 | 25.97 | 25.38 | 26.13 | 26.46 |
| FF | 0.91 | 0.91 | 0.91 | 0.91 | 0.91 | 0.91 |
| PCE (%) | 26.77 | 30.06 | 30.69 | 29.86 | 30.59 | 30.88 |

**Figure 2** shows PCE as a function of cell thickness. The PCE limit of the perovskite cell is about 31%, which approaches to the Shockley-Queisser limit (33%) achievable by gallium arsenide (GaAs) solar cells [29]. Compared to 500 nm-thick flat GaAs solar cells [29] with fill factor (FF) of 0.89, open-circuit voltage ($V_{oc}$) of 1.16 V, and short-circuit current ($J_{sc}$) of 29.5 mA/cm$^2$, perovskite cells have comparable FF, larger $V_{oc}$ and smaller $J_{sc}$. **Table 1** lists the electrical characteristics for $CH_3NH_3PbI_3$ perovskite solar cells with different light-trapping structures. First, both flat cell [See **Fig. 1(a)**] and textured cell [See **Fig. 1(b)**] achieve almost the same FF reaching over 0.9, which is independent of the cell thickness. The very high FF is attributed to drastically suppressed bulk recombination by photon recycling effect. Second, for both solar cells, if the cell thickness increases, $V_{oc}$ slightly decreases. Considering $V_{oc} = \ln(J_p/J_0 + 1)(k_B T/q)$, we know the increase rate of dark current is faster than that of photocurrent. Third, the textured structure boosts $J_{sc}$ compared to the flat structure particularly for thin-film perovskite cells. Although perovskite material has a strong optical absorption, light-trapping design with light-scattering textures still improves the device performance remarkably. For example, a 200 nm-thick textured cell gains the same $J_{sc}$ as a 500 nm-thick flat cell. The textured light-trapping design achieves a very high PCE (29.86%) with just a 200 nm thickness. Using nanostructured plasmonic and photonic designs [37, 38], which break the limit of $4n_r^2$ enhancement [Eq. (4)] at a specific wavelength range, the thickness of perovskite cell can be further reduced for saving material cost.

The above theoretical results assume that the maximum angle of emission $\theta_m$ is 90° without angular restrictions. Importantly, our results show that restricting the angle of light emission will offer a great help to improve the performance of solar cells. It is well known that direct sunlight comes from only a small portion of sky [39]. To convert the direct sunlight to electric power, a solar cell does not need to receive sunlight from all the directions. Therefore, we can design an angular-restriction solar cell [as shown in **Fig. 1(c)**] to only receive the direct (vertically incident) sunlight. With the angular-restriction configuration, $\theta_m$ [in Eqs. (4) and (7)] decreases. As a result, an incident photon with an incident angle larger than $\theta_m$ cannot be absorbed by the solar cell because of detailed balance between emission and absorption. The angular-restriction solar cell has to track the Sun's position and cannot harvest diffused sunlight. However, it can achieve unprecedentedly high $V_{oc}$ and PCE resulting from a strongly reduced dark current $J_0$. **Figure 3** shows the maximum emission angle-dependent PCE for perovskite solar cells. At the wavelengths with (without) the angular-restriction filter, the direct AM 1.5D (global AM 1.5G) spectrum of Sun is employed in the calculations. If a wavelength-independent angular-restriction filtering is implemented for whole effective wavelengths (220 nm--1000 nm), PCE cannot be improved for both flat and textured cells in comparison to the cases without the angular restrictions. It is because that diffused sunlight over all the effective wavelengths cannot be harvested by the solar cell incorporating the wavelength-independent angular-restriction filter. Contrarily, if a wavelength-dependent angular-restriction filter, only covering the wavelengths from 700 nm to 1000 nm, is set up, the PCE is pronouncedly boosted for both cells. At the wavelengths from 700 nm to 1000 nm, the absorption of perovskite materials is very weak (See **Fig. S1**) and thus photocurrent is quite small. However, from 700 nm to 1000 nm, the ratio of dark current $J_0$ over photocurrent $J_p$ is large, compared to that from 280 nm to 700 nm. The sun spectrum can be approximated as a thermal radiation spectrum with a temperature of 5780 K. Differently, the dark current of perovskite cell is related to thermal energy of 300 K. The Sun and perovskite cell have different thermal radiation spectra, which induces a large ratio of dark current over photocurrent ($J_0/J_p$) from 700 nm to 1000 nm. The wavelength-dependent angular-restriction design enables the textured perovskite solar cell with a thickness of 200 nm to achieve the Shockley-Queisser limit (33%), as illustrated in **Fig. 3(b)**.

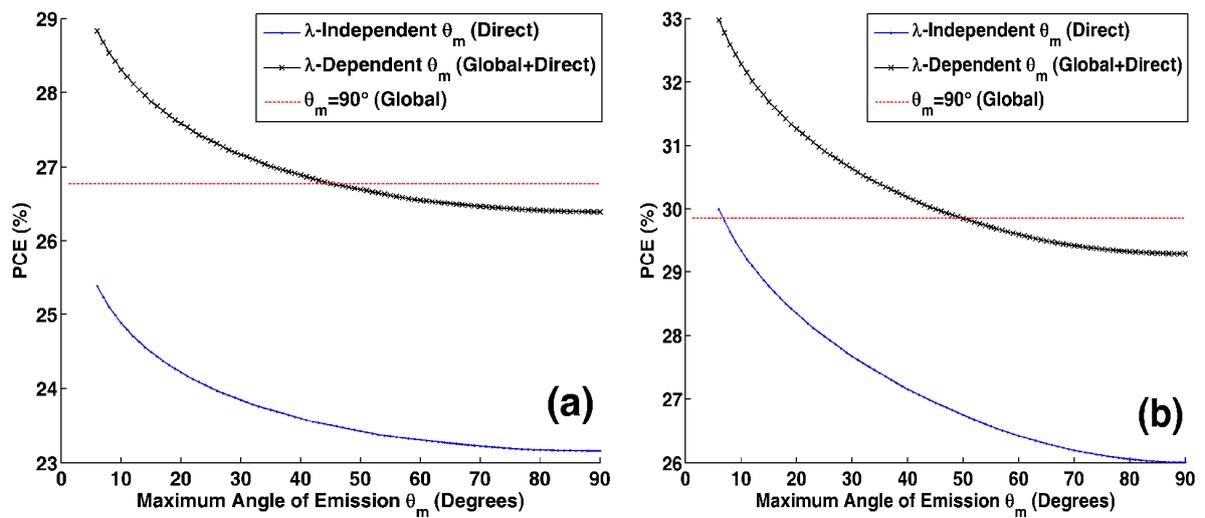

**FIG. 3.** (Color online) PCE of $CH_3NH_3PbI_3$ perovskite solar cells as a function of the maximum angle of emission $\theta_m$. (a) flat cells with a thickness of 200 nm; (b) textured cells with a thickness of 200 nm. Flat and textured cells adopt different light-trapping structures as depicted in **Figs. 1(a) and 1(b)**. For wavelength (λ)-independent angular-restriction designs, an angular-restriction filter is applied to whole effective wavelength range of 220 nm to 1000 nm. For wavelength (λ)-dependent angular-restriction designs, the filter is applied only to the wavelength range of 700 nm to 1000 nm. The results without the angular restrictions ($\theta_m = 90°$) are also given for comparisons. At the wavelengths with (without) the angular-restriction filtering, the direct AM 1.5D (global AM 1.5G) spectrum of Sun is used in the calculations.

Currently, experimentally achievable efficiency of perovskite solar cells is still well below the efficiency limit predicted above. Trap-assisted nonradiative recombination [40] is the dominant mechanism for limiting device performance. After incorporating the trap-assisted monomolecular (or Shockley-Read-Hall) recombination to the detailed balance model [41], the influence of nonradiative recombination on electrical responses of perovskite solar cells is investigated (See **Figs. S2 and S3** in Supplementary Material [36]). First, device efficiencies of both flat and textured cells are significantly reduced by the nonradiative recombination. An optimized cell thickness has to be carefully selected for reaching maximum efficiency. For example, active layer of textured cells should be smaller than 200 nm for achieving the highest PCE. Second, nonradiative recombination will also affect the angular-restriction design. For flat cells, no efficiency improvement is found even if we employed the wavelength-dependent angular restriction. Under open-circuit condition, "radiative" current of flat perovskite cells is ignorable compared to a large nonradiative (recombination) current. The ignorable "radiative" current as described in Eq. (5) is caused by a very low optical absorption of $CH_3NH_3PbI_3$ at long wavelengths, which is different from GaAs material. For textured perovskite cells, the angular-restriction design still increases the efficiency due to the improved absorptivity in Eq. (4). Third, according to experimental data [5,42] and the detailed balance model, the intrinsic carrier density of $CH_3NH_3PbI_3$ material is around $10^7$ cm$^{-3}$, which is much smaller than that of silicon (~$10^9$ cm$^{-3}$) and is larger than that of GaAs (~$10^6$ cm$^{-3}$).

In conclusion, the efficiency limit of $CH_3NH_3PbI_3$ perovskite solar cells is predicted with the detailed balance model. First, the efficiency limit of the perovskite cell (without the angular restriction) is about 31%, which is close to the Shockley-Queisser limit (33%). Second, light-trapping designs still play an important role in improving the efficiency of thin-film perovskite cells, even if perovskite material has a strong optical absorption. Third, a well-engineered angular-restriction design promotes a textured thin-film perovskite cell (200 nm) to reach the Shockley-Queisser limit. All above striking results are based on the assumption of photon recycling effect in perovskite materials, which still needs more experimental studies and evidences.

This project is supported by the University Grant Council of the University of Hong Kong (grants 201311159056), the General Research Fund (grants HKU711813 and HKU711612E), the Collaborative Research Fund (grant C7045-14E) and RGC-NSFC grant (N_HKU709/12) from the Research Grants Council of Hong Kong Special Administrative Region, China, and grant CAS14601 from CAS-Croucher Funding Scheme for Joint Laboratories. This project is also supported by The National Natural Science Foundation of China (No. 61201122) and in part by a Hong Kong UGC Special Equipment Grant (SEG HKU09).

## Supplementary Material

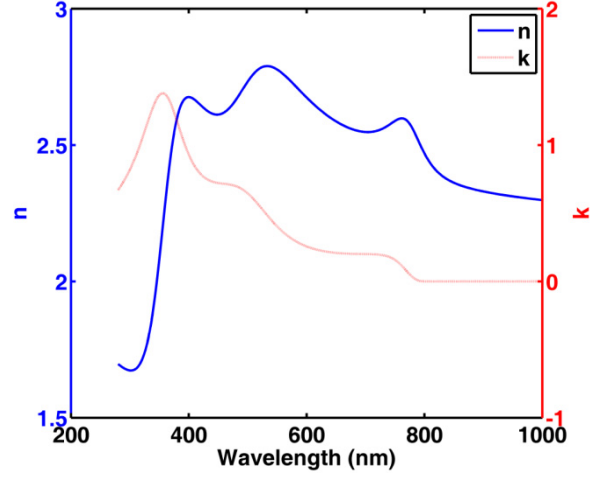

**FIG. S1.** Complex refractive index of CH₃NH₃PbI₃ perovskite materials.

**Nonradiative Recombination Incorporated Detailed Balance Model**

To analyze the influence of trap-assisted nonradiative recombination on the efficiency limit of CH₃NH₃PbI₃ solar cells, the detailed balance model should be modified as

$$J(V) = J_t(V) + J_e(V) - J_p \qquad (S1)$$

where $V$ is the applied voltage of the solar cell system. $J_p$ is the photocurrent (photogenerated current) due to absorption of incident sunlight. $J_e$ and $J_t$ describe current density losses due to the radiative emission and trap-assisted nonradiative recombination, respectively. For perovskite solar cells, the dominant nonradiative recombination process is the monomolecular (or Shockley-Read-Hall) recombination. Therefore, the nonradiative current can be expressed as

$$J_t(V) = q\gamma n_i W \exp\left(\frac{0.5qV}{k_B T}\right) \qquad (S2)$$

where $\gamma$ is the monomolecular (recombination) rate, $n_i$ is the intrinsic carrier density, and $W$ is the cell thickness. $q$ is the elementary charge, $k_B$ is the Boltzmann constant, and $T$ is the kelvin temperature.

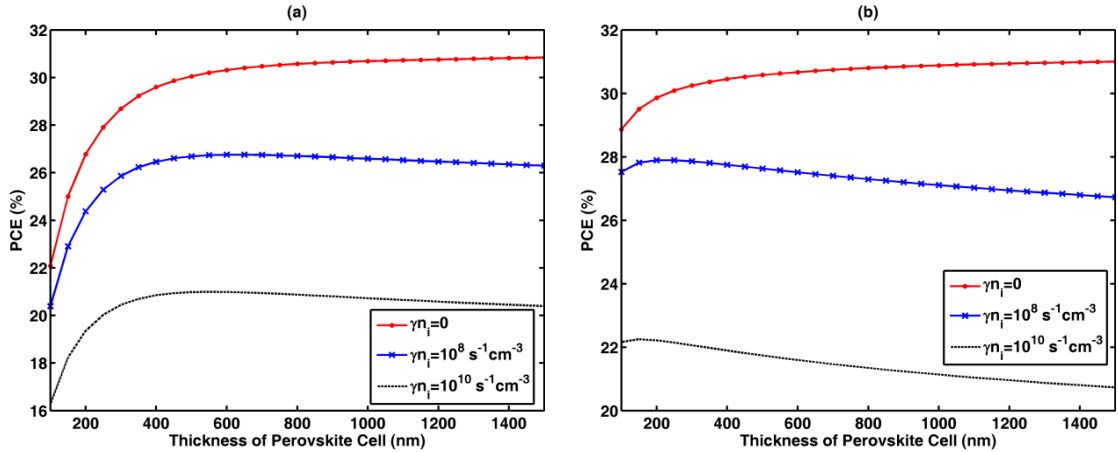

**FIG. S2**. Nonradiative-recombination dependent device efficiency of perovskite solar cells (without angular-restriction designs). The nonradiative recombination is proportional to the product of the monomolecular (recombination) rate $\gamma$ and intrinsic carrier density $n_i$. (a) flat cell with a perfectly reflecting mirror; (b) textured cell with a perfectly reflecting mirror.

**Figure S2** shows nonradiative-recombination dependent device efficiency of perovskite solar cells. For both flat and textured cells, nonradiative recombination degrades the device efficiency significantly. An optimized thickness of active layer should be carefully selected for different light-trapping designs. For example, the active-layer thickness of textured cells should be smaller than 200 nm for achieving the best efficiency. **Figure S3** depicts the influence of nonradiative recombination on the angular-restriction designs. For flat cells, no improvements can be found even if we have adopted the wavelength-dependent angular restrictions. In view of a very weak optical absorption at long wavelengths, the radiative current of flat perovskite cells is quite small compared to the nonradiative (recombination) current, particularly under the maximum power point and open-circuit voltage conditions. This feature is critically different from GaAs materials with a stronger optical absorption at long wavelengths. Thus, the radiative and nonradiative currents of GaAs solar cells have comparable amplitudes. In contrast to flat perovskite cells, the device efficiency of textured perovskite cells still can be improved due to enhanced absorptivity as presented in Eq. (4) of manuscript.

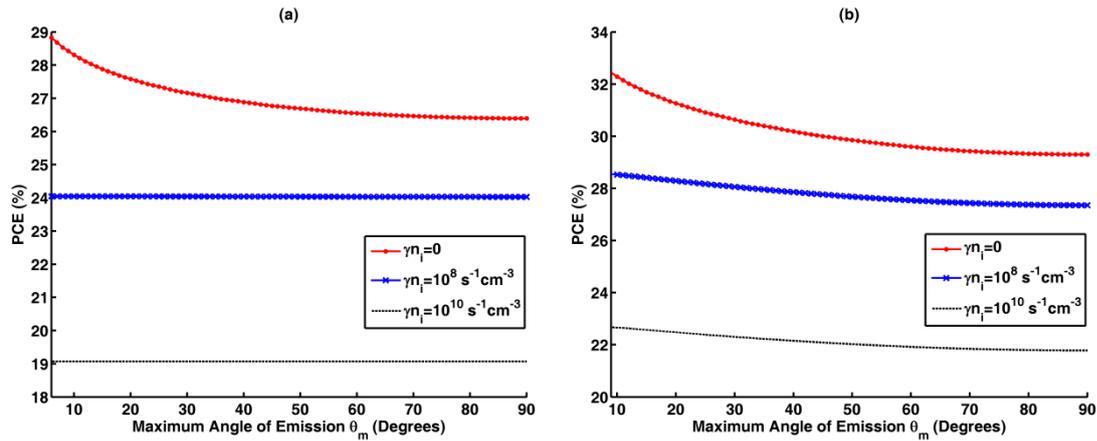

**Figure S3**. Device efficiency of perovskite solar cells as a function of maximum emission angle. The influence of nonradiative recombination on angular-restriction designs is investigated. The nonradiative recombination is proportional to the product of the monomolecular (recombination) rate $\gamma$ and intrinsic carrier density $n_i$. A wavelength-dependent angular-restriction design is adopted, i.e. an angular-restriction filter is applied only to the wavelength range of 700 nm to 1000 nm. At the wavelengths with (without) the angular-restriction filtering, the direct AM 1.5D (global AM 1.5G) spectrum of Sun is used in the calculations. (a) flat cell with a perfectly reflecting mirror; (b) textured cell with a perfectly reflecting mirror.